\documentclass[a4paper,11pt]{article}
\pdfoutput=1

\usepackage{makecell}
\usepackage{orcidlink}
\usepackage{jheppub}

\usepackage[T1]{fontenc} % if needed
\usepackage{float} 
\usepackage{lipsum}
\usepackage{capt-of} 
\usepackage{adjustbox}

\usepackage{graphicx}

\usepackage{lineno}

\title{ \boldmath \quad\\[0.5cm] Search for lepton-flavor-violating tau decays to $\ell\alpha$ at Belle}

\collaboration{The Belle Collaboration}
\collaborationImg{\includegraphics[width=2cm]{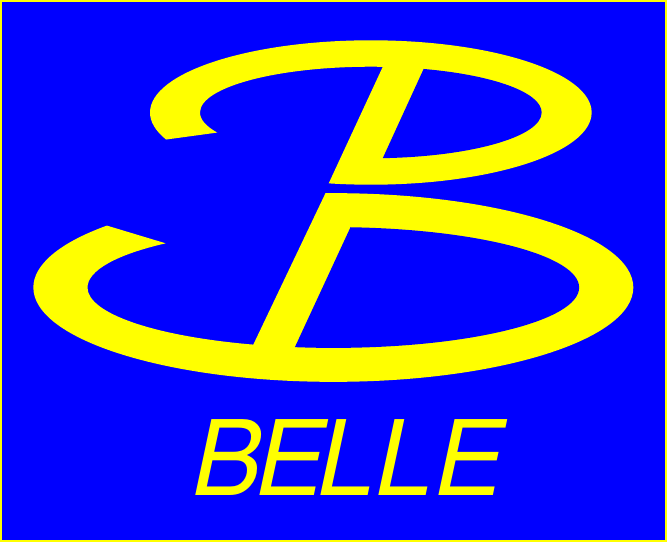}}
 \author{K.~Uno,\orcidlink{0000-0002-2209-8198},} % 14963
 \author{K.~Hayasaka,\orcidlink{0000-0002-6347-433X},} % 2330
 \author{K.~Inami,\orcidlink{0000-0003-2765-7072},} % 2323
  \author{H.~Aihara,\orcidlink{0000-0002-1907-5964},} % 2223
  \author{R.~Ayad,\orcidlink{0000-0003-3466-9290},} % 3766
  \author{Sw.~Banerjee,\orcidlink{0000-0001-8852-2409},} % 8603
  \author{K.~Belous,\orcidlink{0000-0003-0014-2589},} % 2329
  \author{J.~Bennett,\orcidlink{0000-0002-5440-2668},} % 2454
  \author{M.~Bessner,\orcidlink{0000-0003-1776-0439},} % 3783
  \author{D.~Biswas,\orcidlink{0000-0002-7543-3471},} % 8703
% ...
\emailAdd{kenta.uno@kek.jp}

\abstract{
  We report a search for the lepton-flavor-violating decays $\tau^{\pm}\to\ell^{\pm}\alpha$~($\ell=e,\mu$), where $\alpha$ is an undetected spin-0 particle, such as an axion-like particle using $736\times10^{6}$ tau lepton pairs collected by the Belle detector at the KEKB asymmetric-energy $e^{+}e^{-}$ collider.
  We find no evidence of signal and obtain the most stringent upper limits on the branching fractions at 95\% confidence level: $\mathcal{B}(\tau^{\pm}\rightarrow e^{\pm}\alpha)$ $<$ $(0.4$--$6.4)\times10^{-4}$ and $\mathcal{B}(\tau^{\pm}\rightarrow \mu^{\pm}\alpha)$ $<$ $(0.2$--$3.5)\times10^{-4}$ at 95\% confidence level for an $\alpha$ mass in the range $0.0\leq m_{\alpha}\leq 1.6$~GeV/$c^{2}$.
}


\preprint{\vbox{ \hbox{   }
\hbox{Belle Preprint 2025-01}
\hbox{KEK Preprint 2025-3}
}}

\begin{document} 
\maketitle
\flushbottom
\section{Introduction}

The Standard Model~(SM)~\cite{SM} has been successful in providing a description of particle physics.
Yet new models beyond the Standard Model~(BSM) are proposed and discussed in order to solve problems that cannot be explained by the SM.
Among such problems are the particle mass hierarchy, the neutrino mass, and the existence of dark matter~\cite{Darkmatter}.
Many BSM models explain these phenomena by introducing new long-lived bosons~\cite{ALP_NP1,ALP_NP2,ALP_NP3}.
For instance, models explaining dark matter or the muon magnetic-moment anomaly can do so through the existence of axion-like particles~(ALP) which would have spin-0 and interact feebly with ordinary matter.

Experiments such as MARKIII, ARGUS and Belle II~\cite{MARK2,ARGUS,BelleII} have searched for such a particle, here called $\alpha$, by studying the lepton-flavor-violating decay $\tau^{-}\to\ell^{-}\alpha$\footnote{Throughout this paper, the inclusion of charge-conjugate decay modes is implied. }, where $\ell=e,\mu$ and $\alpha$ is assumed to fly undetected.
The most stringent upper limits on the branching fractions are provided by the Belle II experiment using a data sample of 62.8~fb$^{-1}$~\cite{BelleII}:
$\mathcal{B}(\tau^{-}\to e^{-}\alpha) < (0.19-1.73)\times10^{-3}$ and $\mathcal{B}(\tau^{-}\to\mu^{-}\alpha) < (0.13-2.13)\times10^{-3}$ at the 95\% confidence level~(CL) for an $\alpha$ mass in the range $0.0\leq m_{\alpha}\leq 1.6$~GeV/$c^{2}$.

In this paper, a search for $\tau^{-}\rightarrow\ell^{-}\alpha$ decays at the Belle experiment is reported.
We use data recorded at the $\Upsilon(4S)$ resonance corresponding to a luminosity of 711~fb$^{-1}$. In addition, a data sample of 89~fb$^{-1}$ recorded 60~MeV below the $\Upsilon(4S)$ resonance is used~\cite{Luminosity}.
The total integrated luminosity is 800 fb$^{-1}$, which corresponds to $736\times10^{6}$ tau pairs~($N_{\tau\tau}$)~\cite{Ntautau}.

The Belle detector is a large-solid-angle magnetic spectrometer consisting of a silicon vertex detector~(SVD), a 50-layer central drift chamber~(CDC), an array of aerogel threshold Cherenkov counters~(ACC), a barrel-like arrangement of time-of-flight scintillation counters~(TOF), and an electromagnetic calorimeter composed of CsI(Tl) crystals~(ECL). These components are located inside a superconducting solenoid coil that provides a 1.5~T magnetic field at the asymmetric-energy $e^{+}$~(3.5~GeV)~$e^{-}$~(8~GeV) KEKB collider~\cite{KEKB0,KEKB}.  An iron flux-return located outside of the coil is instrumented with resistive plate chambers to detect $K_L^0$ mesons and muons~(KLM). The $z$-axis of the detector points in the direction opposite to the positron beam.
The detector is described in detail elsewhere~\cite{Luminosity,Belle}. 

We use Monte Carlo~(MC) simulated samples to optimize the event selection, to calculate the signal efficiency, and to estimate the signal and the background contributions.
Signal MC samples and SM tau decays are generated using the KKMC package and TAUOLA~\cite{KKMC}.
Other background processes, namely, $e^{+}e^{-}\to e^{+}e^{-}\gamma$~($e^{+}e^{-}\gamma$), $e^{+}e^{-}\to\mu^{+}\mu^{-}\gamma$~($\mu^{+}\mu^{-}\gamma$), $e^{+}e^{-}\rightarrow e^{+}e^{-}\ell^{+}\ell^{-}$~(two-photon), and $e^{+}e^{-}\rightarrow q\bar{q}, q=u,d,s,c,b~(q\bar{q})$ events are generated using BHLUMI~\cite{BHLUMI}, KKMC~\cite{KKMC}, AAFH~\cite{AAFHB}, and EvtGen~\cite{EvtGen}, respectively.
Signal MC samples are $e^{+}e^{-}\to\tau^{+}\tau^{-}$~($\tau^{+}\tau^{-}$) events with one of the taus decaying to the $\ell^{\pm}\alpha$ final state and the other according to the known branching fractions~\cite{PDG}.
We generate signal MC samples with $\alpha$ masses~($m_{\alpha}$) from 0 up to $1.6$ GeV/$c^{2}$. The detector simulation is performed using GEANT3~\cite{GEANT3}.

\section{Reconstruction and event selection}
A $\tau^{+}\tau^{-}$ event is divided into two hemispheres in the center-of-mass~(c.m.) frame using a thrust vector $\hat{t}$~\cite{thrust}.
The vector $\hat{t}$ is defined so that the magnitude of thrust in the event, $T = \sum \frac{ |\vec{p}_{i}^{\mathrm{~c.m.}}\cdot\hat{t}| }{\sum |\vec{p}_{i}^{\mathrm{~c.m.}}|}$ is maximized.
Here, $\vec{p}_{i}^{~\mathrm{c.m.}}$ is the three-momentum of the $i$-th final-state particle in the c.m. frame and the sum runs over all charged and neutral particles.
We assign every particle to either the signal-side tau~($\tau_{\mathrm{sig}}$) or tag-side tau~($\tau_{\mathrm{tag}}$), respectively.
The signal-side tau is expected to decay to $\tau^{-}_{\mathrm{sig}}\to\ell^{-}\alpha$ and the tag-side tau is required to be reconstructed in a one-prong or three-prong decay such as $\tau^{+}_{\mathrm{tag}}\to\ell^{+}\nu_{\ell}\bar{\nu}_{\tau}$, $\tau^{+}_{\mathrm{tag}}\to h^{+}\bar{\nu}_{\tau}$ and $\tau^{+}_{\mathrm{tag}}\to h^{+}h^{-}h^{+}\bar{\nu}_{\tau}$~($h=\pi,K$).
If the number of tracks in the tag hemisphere is one, the event is assigned to the one-prong channel. Otherwise, the event is assigned to the three-prong channel.

\begin{figure}[htbp]
 \begin{minipage}{0.45\hsize}
  \begin{center}
   \includegraphics[width=70mm]{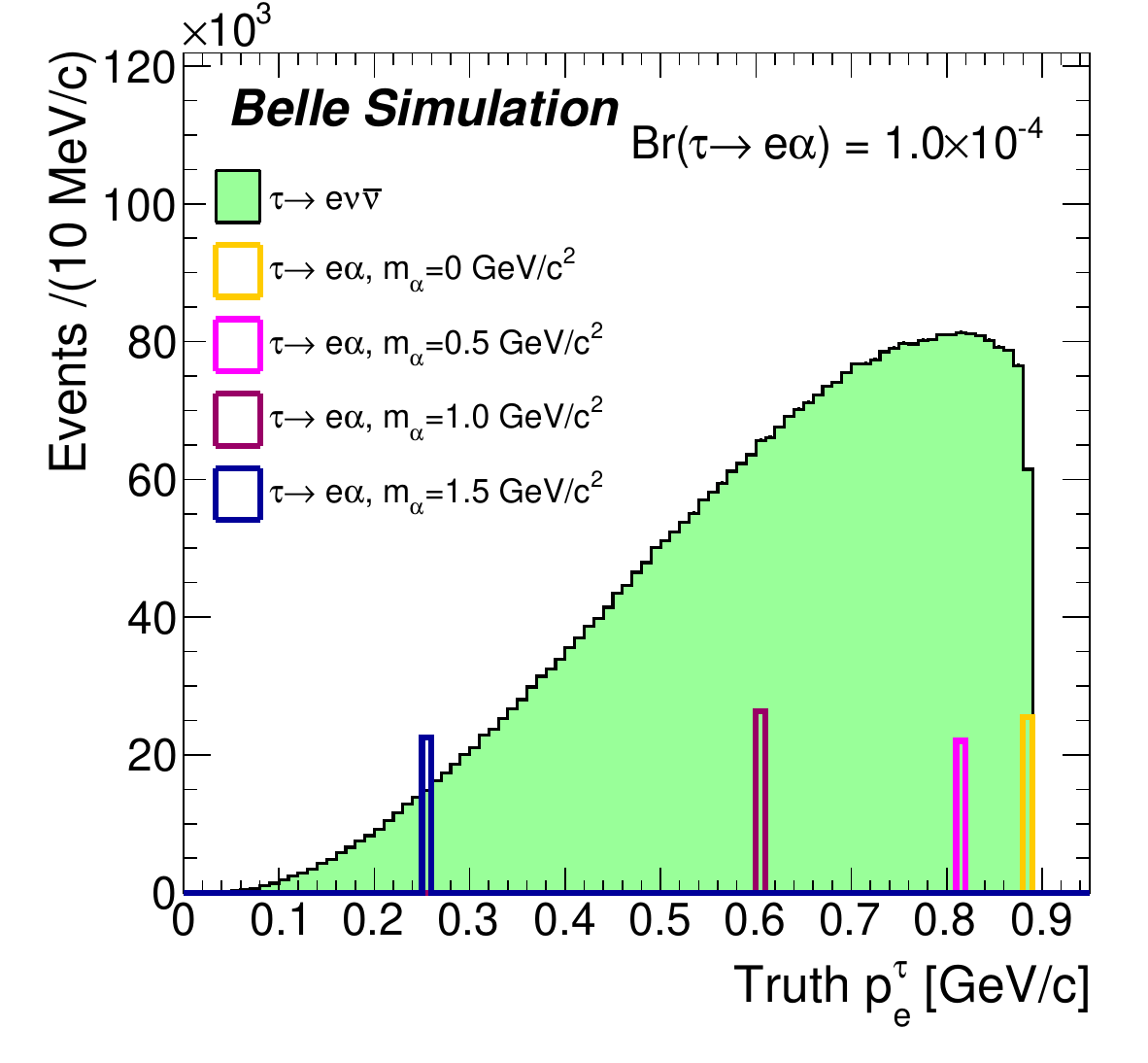}
  \end{center}
 \end{minipage}
 \begin{minipage}{0.45\hsize}
  \begin{center}
    \includegraphics[width=70mm]{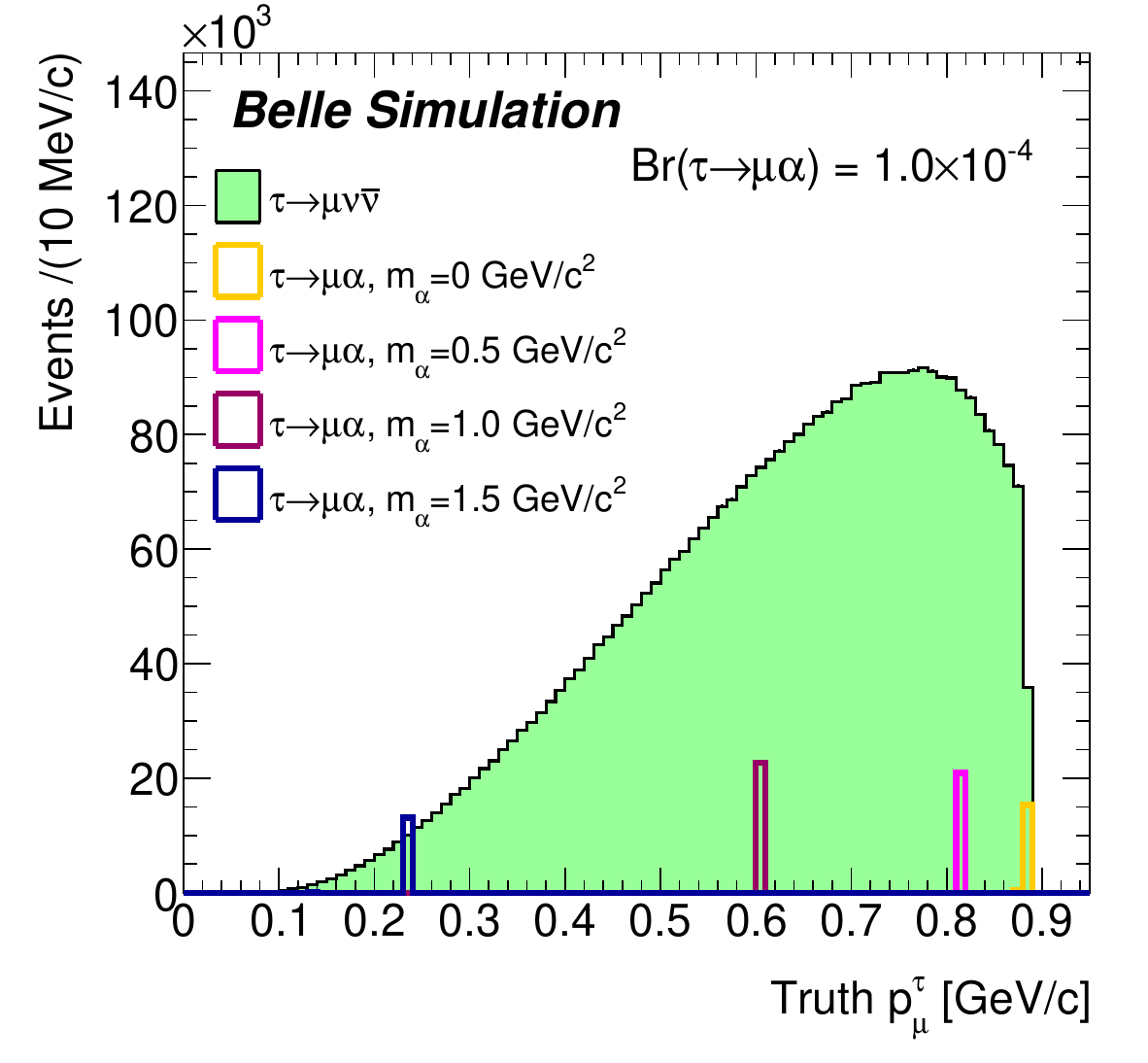}
  \end{center}
 \end{minipage}
\caption{ \label{fig:truth_pltau} 
  Distributions of the truth $p_{\ell}^{\tau}$ for $\tau^{-}\rightarrow e^{-}\alpha$ and $\tau^{-}\rightarrow e^{-}\bar{\nu}_{e}\nu_{\tau}$ events~(left) and $\tau^{-}\rightarrow \mu^{-}\alpha$ and $\tau^{-}\rightarrow \mu^{-}\bar{\nu}_{\mu}\nu_{\tau}$ events~(right).
  The signal MC distributions are shown for an arbitrary branching fraction $\mathcal{B}(\tau^{-}\to\ell^{-}\alpha)=1.0\times10^{-4}$.
}
\end{figure}

The signal tau decays to a two-body final state, while the dominant background events are found to originate from the three-body decays, $\tau^{-}_{\mathrm{sig}}\to\ell^{-}\bar{\nu}_{\ell}\nu_{\tau}$.
The key variable to separate signal and background events is the lepton momentum in the tau rest frame $p_{\ell}^{\tau}$.
Figure~\ref{fig:truth_pltau} shows the distribution of truth $p_{\ell}^{\tau}$ for simulated $\tau^{-}\rightarrow \ell^{-}\alpha$ and $\tau^{-}\rightarrow \ell^{-}\bar{\nu}_{\ell}\nu_{\tau}$ events.
The distribution for the signal events is monochromatic and is dependent on the $\alpha$ mass, whereas the $\tau^{-}\to\ell^{-}\bar{\nu}_{\ell}\nu_{\tau}$ events have a broad distribution. By utilizing this signature, we can distinguish signal events from the $\tau^{-}\to\ell^{-}\bar{\nu}_{\ell}\nu_{\tau}$ background.

\begin{figure}[htbp]
 \begin{minipage}{0.45\hsize}
  \begin{center}
   \includegraphics[width=70mm]{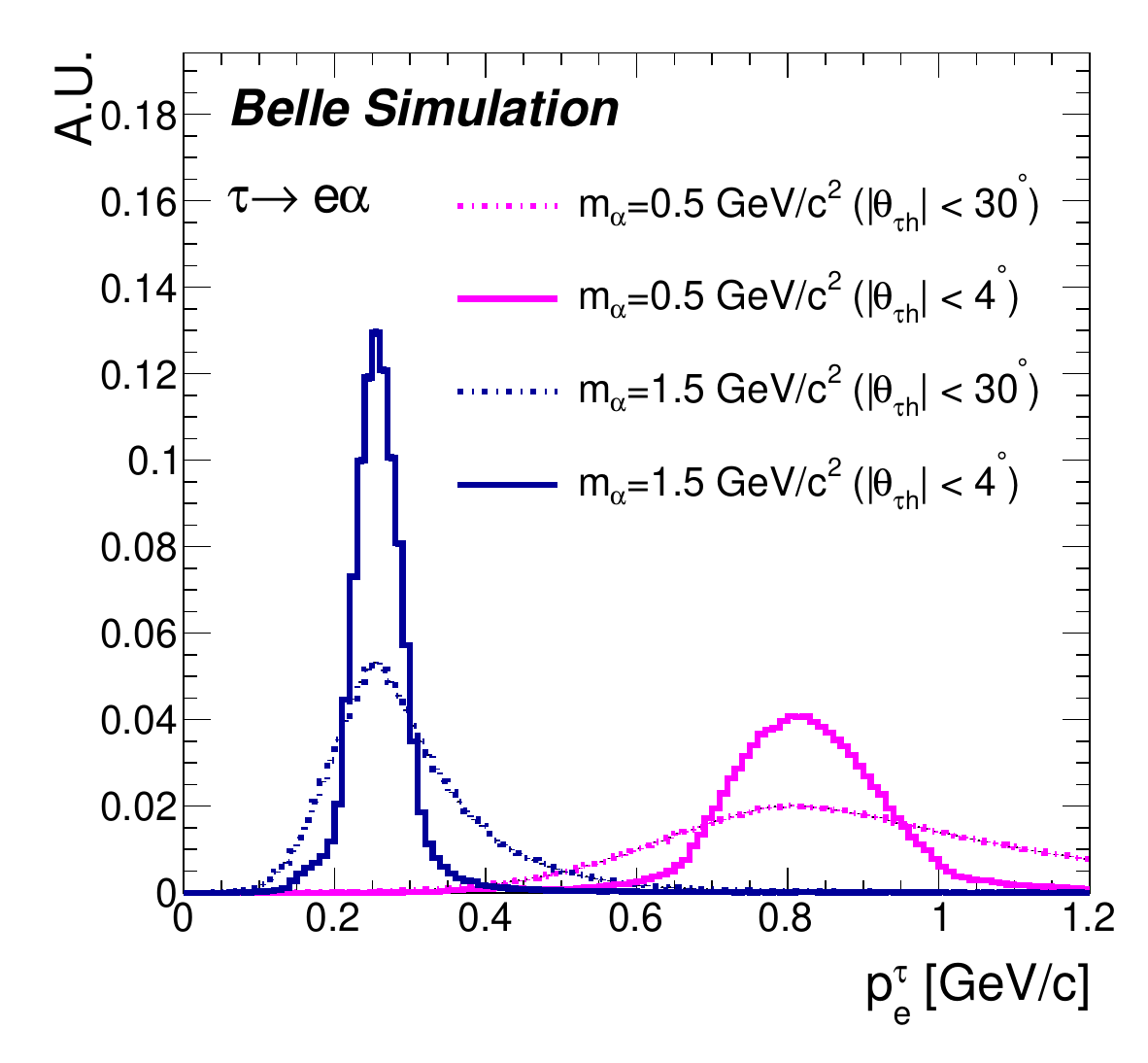}
  \end{center}
 \end{minipage}
 \begin{minipage}{0.45\hsize}
  \begin{center}
    \includegraphics[width=70mm]{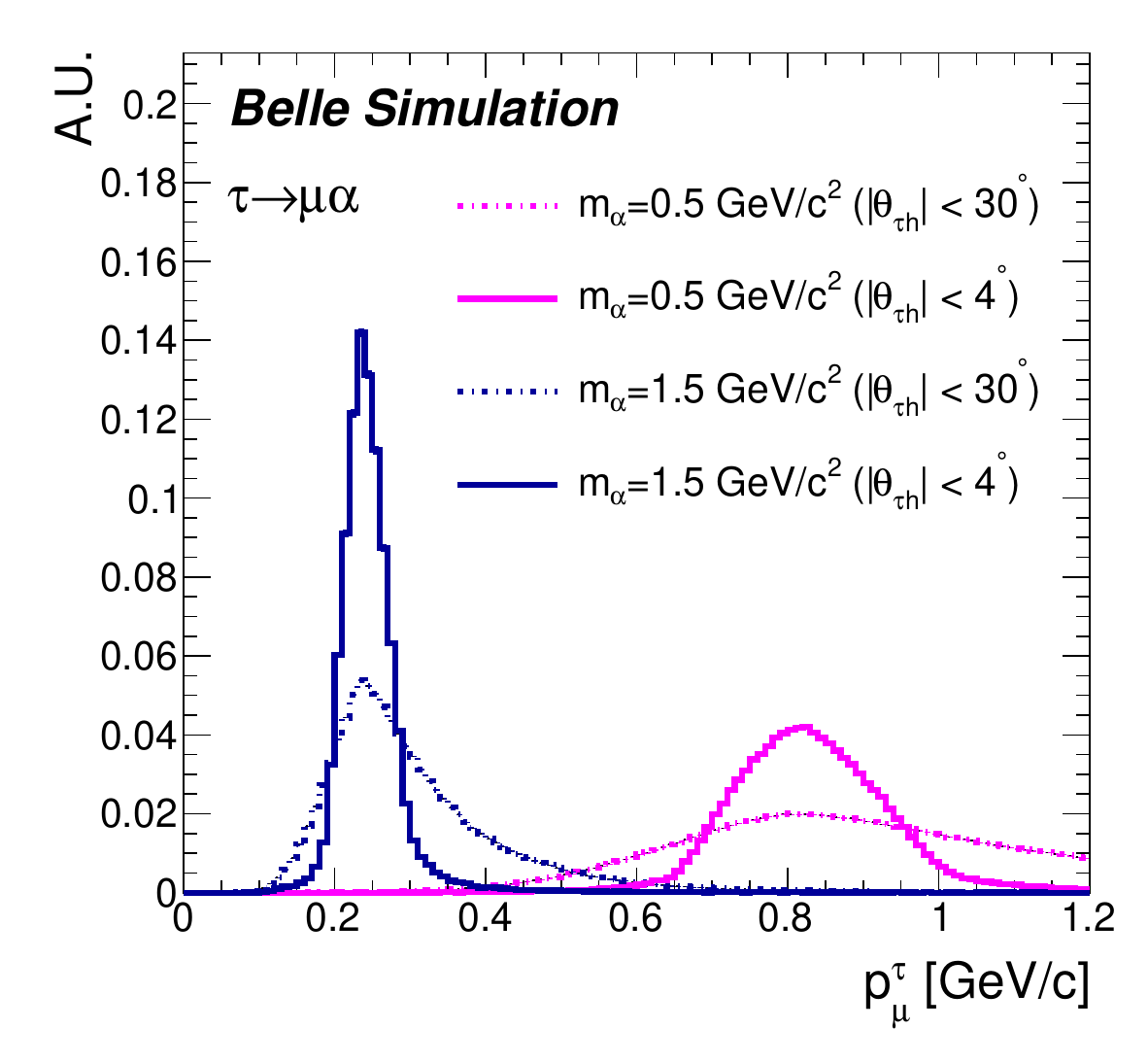}
  \end{center}
 \end{minipage}
 \caption{ \label{fig:reco_pltau_compare} 
   Distributions of $p_{\ell}^{\tau}$ with different $\theta_{\tau h}$ selections are shown for simulated signal events with $m_{\alpha}=500,1500$~MeV/$c^{2}$.
   The left figure shows the distributions for $\tau^{-}\to e^{-}\alpha$ events and the right figure shows the distributions for $\tau^{-}\to \mu^{-}\alpha$ events.
   The solid~(dashed) lines show the distribution with $|\theta_{\tau h}| < 4^{\circ}$~($|\theta_{\tau h}| < 30^{\circ}$).
}
\end{figure}

In order to obtain $p_{\ell}^{\tau}$, the momentum and the flight direction of the tau lepton are needed.
Due to the $\alpha$ being invisible to the Belle detector, the flight direction of the tau lepton cannot be determined. In this analysis, the following approximation is used.
For the tag-side hadronic tau decays, we compute the opening angle~($\theta_{\tau h}$) between the hadronic particles $h_{\mathrm{tag}}$ and $\tau^{+}_{\mathrm{tag}}$.
Here, $h_{\mathrm{tag}}$ includes a combination of charged and neutral particles like $\pi^{\pm}\pi^{0}$.
The four-momentum of the neutrino in the tag side, $p_{\nu_{\mathrm{tag}}}^{\mathrm{~c.m.}}=(E_{\nu_{\mathrm{tag}}}^{\mathrm{~c.m.}},~\vec{p}_{\nu_{\mathrm{tag}}}^{\mathrm{~c.m.}})$, can be expressed as
\begin{eqnarray} \label{eq:pnu_0}
  (p_{\nu_{\mathrm{tag}}}^{\mathrm{~c.m.}})^{2} &=& (p_{\tau_{\mathrm{tag}}}^{\mathrm{~c.m.}} - p_{h_{\mathrm{tag}}}^{\mathrm{~c.m.}})^{2} \nonumber \\ 
  &=& \left(
\frac{\sqrt{s}}{2} - E_{h_{\mathrm{tag}}}^{\mathrm{~c.m.}}
\right)^{2} - |\vec{p}_{\tau_{\mathrm{tag}}}^{\mathrm{~c.m.}}|^{2} - |\vec{p}_{h_{\mathrm{tag}}}^{\mathrm{~c.m.}}|^{2} + 2|\vec{p}_{\tau_{\mathrm{tag}}}^{\mathrm{~c.m.}}||\vec{p}_{h_{\mathrm{tag}}}^{\mathrm{~c.m.}}|\cos\theta_{\tau h},
\end{eqnarray}
where $p_{\tau_{\mathrm{tag}}}^{\mathrm{~c.m.}}$ and $p_{h_{\mathrm{tag}}}^{\mathrm{~c.m.}}$ are the four-momenta of the tau lepton and the hadronic system on the tag side.
Here, we approximate the energy of $\tau^{+}_{\mathrm{tag}}$ in the c.m. frame as $\sqrt{s}/2$ and assume that the neutrino is massless.
The angle $\theta_{\tau h}$ is then given by
\begin{equation} \label{eq:fom4}
  \theta_{\tau h} = \arccos \left(\frac{|\vec{p}_{\tau_{\mathrm{tag}}}^{\mathrm{~c.m.}}|^{2} + |\vec{p}_{h_{\mathrm{tag}}}^{\mathrm{~c.m.}}|^{2} - (\sqrt{s}/2 - E_{h_{\mathrm{tag}}}^{\mathrm{~c.m.}})^{2}}{2|\vec{p}_{\tau_{\mathrm{tag}}}^{\mathrm{~c.m.}}||\vec{p}_{h_{\mathrm{tag}}}^{\mathrm{~c.m.}}|} \right).
\end{equation}
Here, the magnitude of the momentum of $\tau^{+}_{\mathrm{tag}}$ is taken as $|\vec{p}_{\tau_{\mathrm{tag}}}^{\mathrm{~c.m.}}|=\sqrt{(\sqrt{s}/2 - m_{\tau})^{2}}$ where $m_{\tau} = 1776.86\pm 0.12$~MeV/$c^{2}$ is the mass of the tau lepton~\cite{PDG}.
All tag-side tracks are assumed to be charged pion. For the tag-side leptonic tau decays, we compute an opening angle in the same way. 
By requiring a stringent selection on $\theta_{\tau h}$ so that the signal candidates are well aligned with the direction opposite to the tag-side hadronic particle system, we achieve an improvement in the $p_{\ell}^{\tau}$ resolution as shown in Figure~\ref{fig:reco_pltau_compare}. Here, we use the flight direction of the signal-side tau lepton as opposite direction of $h_{\mathrm{tag}}$ in the calculation of $p_{\ell}^{\tau}$.
Since there are two neutrinos for the leptonic tau decays, the angle $\theta_{\tau h}$ tends to be large. Thus, we consider only the hadronic tau decays by requiring the stringent selection.
In order to maximize the search sensitivity, we select signal candidates that satisfy $|\theta_{\tau h}|<4^{\circ}$.
We keep the data in the distribution of $p_{\ell}^{\tau}$ hidden until the selection criteria and background estimation strategy are finalized.

In order to select tau-pair events, two or four charged particles with zero net charge are required.
The reconstructed trajectories~(tracks) are required to have a distance of closest approach to the interaction point~(IP) smaller than 5~mm in the transverse plane $|dr|$ and 50~mm along the $z$ axis $|dz|$.
Candidate events are retained if all tracks satisfy $p^{\mathrm{c.m.}}\geq0.1$ GeV/$c$ and $-0.866<\cos\theta_{\mathrm{track}}<0.956$ in order to reduce background from $e^{+}e^{-}\gamma$, $\mu^{+}\mu^{-}\gamma$, and two-photon events, where $\theta_{\mathrm{track}}$ is the polar angle of the track in the laboratory frame.

The track on the signal side must be identified as an electron or a muon by the particle identification algorithm~\cite{Electron,Muon}.
Identification of electrons uses a likelihood ratio, $\mathcal{L}_{e}$, based on specific ionization from the CDC, the ratio of the energy deposited in the ECL to the momentum measured by the CDC and SVD,
the shower shape in the ECL, hit information from the ACC, and matching between the track and the ECL cluster.
Electron candidates are required to lie within the ECL acceptance, $\cos\theta_{\mathrm{track}}\in[-0.866,-0.652]\cup[-0.602,0.829]\cup[0.854,0.956]$ to suppress background from misidentified hadrons.
The electron identification efficiency for the selection applied $\mathcal{L}_{e}>0.9$ is 80--91\% depending on the $\alpha$ mass, with a pion misidentification probability of 0.07\%~\cite{Electron}.
Muon candidates are identified using a likelihood ratio, $\mathcal{L_{\mu}}$, which is based on the difference between the range of the track calculated from the particle momentum and that measured in the KLM.
The ratio includes the value of $\chi^{2}$ formed from the KLM hit locations with respect to the extrapolated track.
The muon identification efficiency for the selection applied $\mathcal{L_{\mu}}>0.95$ is 80--88\% depending on the $\alpha$ mass, with a pion misidentification probability of 1.0\% in the region $p>1.0$~GeV/$c$~\cite{Muon}.
To suppress the effect of beam-induced backgrounds, we reject events with reconstructed neutral pions and photons on the signal side.
Photons are required to have an energy $E_{\gamma}>0.1$~GeV within the region $-0.625<\cos\theta_{\mathrm{\gamma}}<0.846$, where $\theta_{\gamma}$ is the polar angle of the photon in the laboratory frame.
Neutral pions are reconstructed as photon pairs within the mass region $120<m_{\gamma\gamma}<135$ MeV/$c^{2}$.
On the tag side, the number of photons is required to be less than four for the one-prong channel and less than two for the three-prong channel to allow for tag-side tau decays including neutral pions.

All selection criteria are chosen in order to maximize the search sensitivity, $\epsilon_{\mathrm{sig}}/\sqrt{N_{\mathrm{bkg}}}$, where $\epsilon_{\mathrm{sig}}$ is the overall signal efficiency and $N_{\mathrm{bkg}}$ is the expected number of background events obtained from MC simulation.
The selection criteria are independent of the $\alpha$ mass.
The total visible energy in the c.m. frame, $E_{\mathrm{total}}^{\mathrm{c.m.}}$, is required to be $2.0$~GeV $<E_{\mathrm{total}}^{\mathrm{c.m.}}<10.0$~GeV to reduce background from $e^{+}e^{-}\gamma$ and $\mu^{+}\mu^{-}\gamma$ events.
To suppress the $q\bar{q}$ events, we use the magnitude of the thrust in the event $T$. For the $\tau^{-}\to e^{-}\alpha$~($\tau^{-}\to \mu^{-}\alpha$) search, we require $T>0.82$~($T>0.87$).
The missing three-momentum is calculated by subtracting the sum of the momenta of all tracks and photons from the sum of the beam momenta in laboratory frame.
To suppress background from low-multiplicity events, a requirement on the cosine of the polar angle of the missing three-momentum, $\cos{\theta_{\mathrm{miss}}}$, is optimized separately for each channel.
For the one-prong channel, we require $-0.880 < \cos\theta_{\mathrm{miss}}<0.950$~($-0.870 < \cos\theta_{\mathrm{miss}}<0.970$) in the $\tau^{-}\to e^{-}\alpha$~($\tau^{-}\to \mu^{-}\alpha$) search.
For the three-prong channel, we require $-0.900 < \cos\theta_{\mathrm{miss}}<0.990$~($-0.980 < \cos\theta_{\mathrm{miss}}<0.990$) in the $\tau^{-}\to e^{-}\alpha$~($\tau^{-}\to \mu^{-}\alpha$) search.
The absolute value of the vector sum of the transverse momenta of all measured particles $|\vec{p}_{\mathrm{T,(total)}}|$ is required to be greater than 0.1~GeV/$c$ in the $\tau^{-}\to e^{-}\alpha$ search and 0.4~GeV/$c$ in the $\tau^{-}\to \mu^{-}\alpha$ search.
Similarly, the absolute value of the vector sum of the transverse momenta of both signal-side and tag-side tracks for the one-prong channel $|\vec{p}_{\mathrm{T,(sig,tag)}}|$ is required to be greater than 0.2~GeV/$c$.
These criteria are used to suppress the $e^{+}e^{-}\gamma$ and $\mu^{+}\mu^{-}\gamma$ backgrounds.
In order to further suppress $e^{+}e^{-}\gamma$ and $\mu^{+}\mu^{-}\gamma$ events, we require $p_{\ell}^{\mathrm{c.m.}}<5.0$~GeV/$c$~($p_{\ell}^{\mathrm{c.m.}}<4.8$~GeV/$c$) for the one-prong channel on the $\tau^{-}\to e^{-}\alpha$~($\tau^{-}\to \mu^{-}\alpha$) search.

We use the control channel $\tau^{-}\to\pi^{-}\pi^{0}\nu_{\tau}$ to study possible biases in the selection criteria.
We require one charged pion and one neutral pion on the signal side, and calculate $p_{\ell}^{\tau}$ using the charged pion momentum after applying the selection criteria.
We find good agreement between data and simulation up to $p_{\ell}^{\tau}<0.95$~GeV/$c$ and require signal candidates to be in this region.

\begin{figure}[htbp]
 \begin{minipage}{0.45\hsize}
  \begin{center}
   \includegraphics[width=70mm]{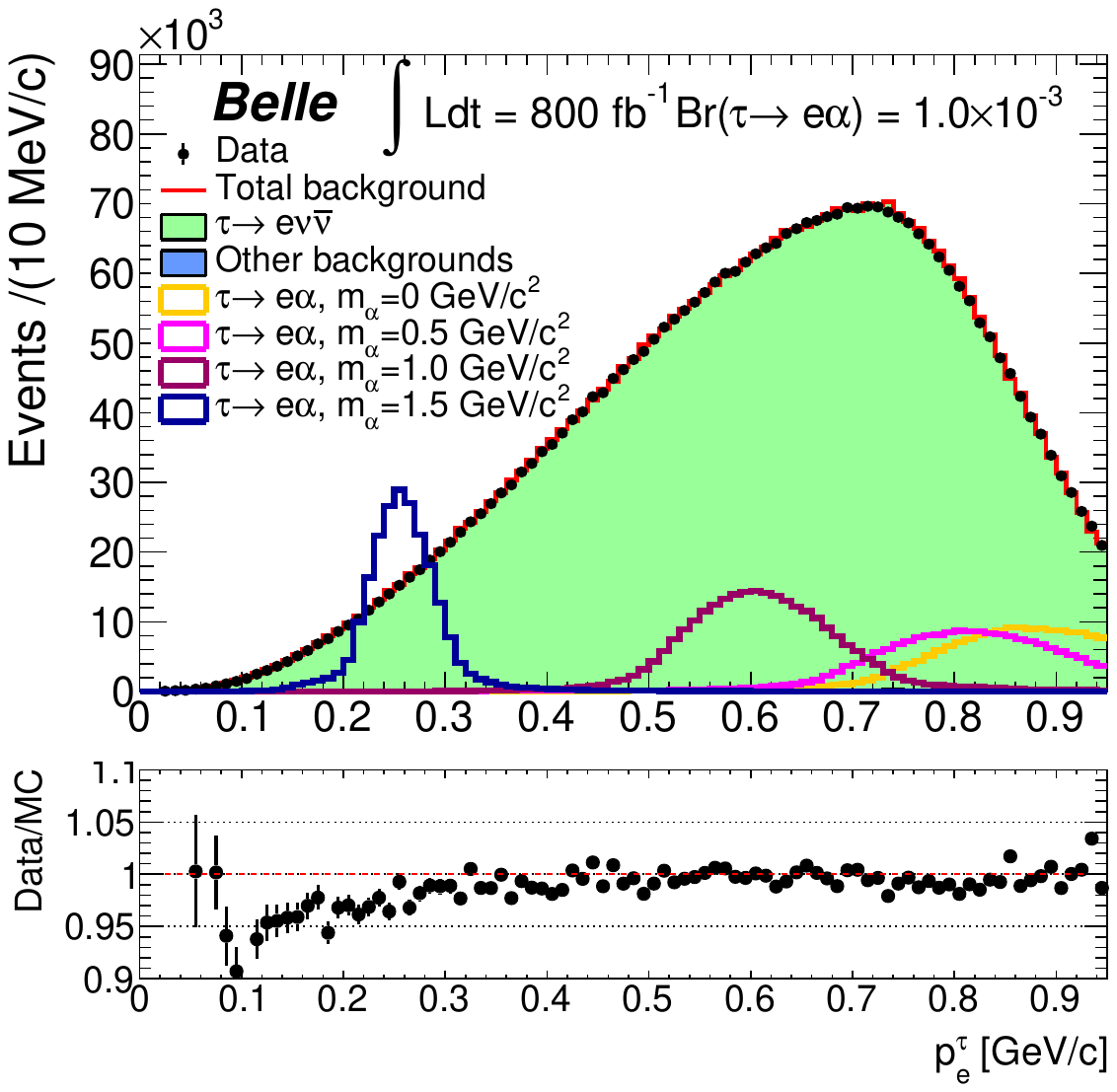}
  \end{center}
 \end{minipage}
 \begin{minipage}{0.45\hsize}
  \begin{center}
    \includegraphics[width=70mm]{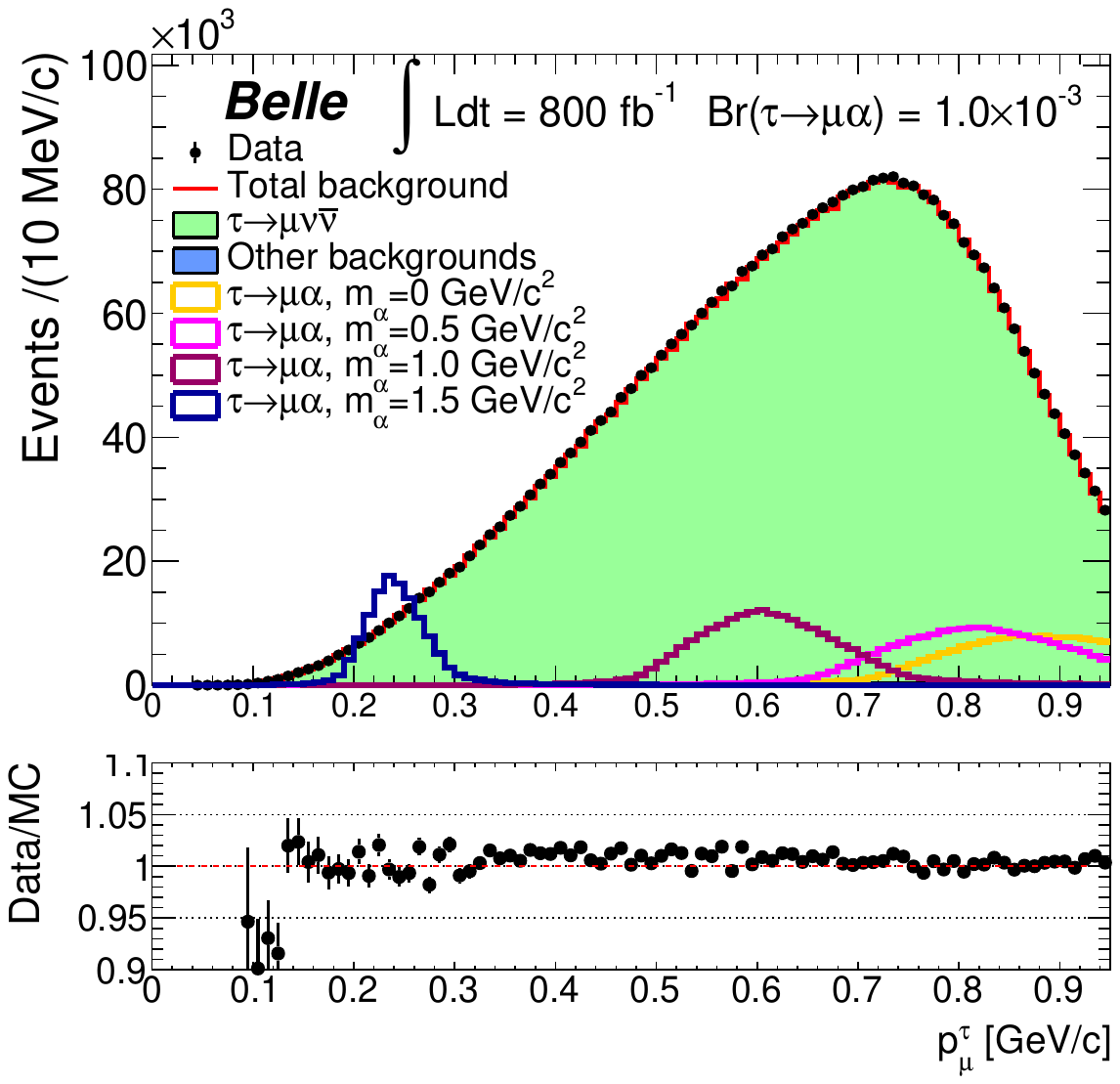}
  \end{center}
 \end{minipage}
\caption{ \label{fig:reco_pltau} 
  Distributions of $p_{\ell}^{\tau}$ for events satisfying all selection criteria are shown.
  The left figure shows the distributions for the $\tau^{-}\to e^{-}\alpha$ search and the right figure shows the distributions for the $\tau^{-}\to \mu^{-}\alpha$ search.
  The background MC samples are normalized to the SM expectation.
  The signal MC distributions are shown for an arbitrary branching fraction $\mathcal{B}(\tau^{-}\to\ell^{-}\alpha)=1.0\times10^{-3}$.
  The bottom figure shows the ratio of the data and the sum of backgrounds.
}
\end{figure}

Figure~\ref{fig:reco_pltau} shows the distributions of $p_{\ell}^{\tau}$ for the selected $\tau\to\ell\alpha$ signal candidates after applying the selection criteria.
The detection efficiencies are $0.9$--$1.4\%$ for the $\tau^{-}\to e^{-}\alpha$ events and $0.3$--$1.5\%$ for $\tau^{-}\to\mu^{-}\alpha$ events, depending on the $\alpha$ mass. The main reason for the efficiency variation is the particle-identification efficiency dependence on the signal-side lepton momentum.
The number of selected events is 3390311 for the $\tau^{-}\to e^{-}\alpha$ search and 3757974 for the $\tau^{-}\to\mu^{-}\alpha$ search.
The dominant background arises from $\tau^{+}\tau^{-}$ events decaying to $\tau^{-}\to\ell^{-}\bar{\nu}_{\ell}\nu_{\tau}$ and contributes 99.2\%~(98.3\%) for the $\tau^{-}\to e^{-}\alpha$~($\tau^{-}\to\mu^{-}\alpha$) search.
Backgrounds from $\tau^{-}\to\pi^{-}\nu_{\tau}$ and $\tau^{-}\to\rho^{-}\nu_{\tau}$ decays contribute 0.2\%~(1.3\%) for the $\tau^{-}\to e^{-}\alpha$~($\tau^{-}\to\mu^{-}\alpha$) search.
The $e^{+}e^{-}\gamma$ and $\mu^{+}\mu^{-}\gamma$ events contribute 0.5\%~(0.4\%) for the $\tau^{-}\to e^{-}\alpha$~($\tau^{-}\to\mu^{-}\alpha$) search.
Other backgrounds including two-photon and $q\bar{q}$ are negligible.

\section{Branching fraction estimation and systematic uncertainty}
We search for narrow peaks in the $p_{\ell}^{\tau}$ distributions by performing binned extended maximum-likelihood fits.
The fits are performed to the selected events for each $\alpha$ mass point and final state in a signal region defined by 0.0~GeV/$c<p_{\ell}^{\tau}<0.95$~GeV/$c$. 17 mass points are considered for each final state.
Both signal and background probability density functions~(PDFs) are obtained by smoothing the corresponding MC distributions. We consider remaining background events after applying the selection criteria such as $\tau^{-}\to\ell^{-}\bar{\nu}_{\ell}\nu_{\tau},\tau^{-}\to\pi^{-}\nu_{\tau},\tau^{-}\to\rho^{-}\nu_{\tau}$, $\mu^{+}\mu^{-}\gamma$ and $e^{+}e^{-}\gamma$ to build the background PDF.
The signal and total background yields are floating parameters.
We perform the likelihood fits to data and obtain the number of observed events~($n_{\mathrm{sig}}$) for each $\alpha$ mass point and final state.
We observe no significant excess of signal events in data over background.
For instance, the number of observed events with $m_{\alpha}=0.0$~GeV/c$^{2}$ are $n_{\mathrm{sig}}=2260.3\pm852.2$ for the $\tau^{-}\to e^{-}\alpha$ search and $n_{\mathrm{sig}}=-764.3\pm723.3$ for the $\tau^{-}\to \mu^{-}\alpha$ search.
We provide an appendix containing the fitted signal yields.

Since no significant excess of signal events is observed in data, we determine 95\% CL upper limits on the branching fractions of $\tau^{-}\to\ell^{-}\alpha$ using a frequentist method, based on the Neyman construction~\cite{ULmethod,Neyman}.

In this method, we generate 10,000 pseudoexperiments with signal and background events based on their PDFs for different signal yields.
We then define the upper limit of the signal yield at 95\% CL~($n_{\mathrm{sig}}^{95}$) as the generated signal yield for which 5\% of the experiments have fitted signal yields less than $n_{\mathrm{sig}}$ in data.
The procedure is performed for each $\alpha$ mass point and final state.
The upper limits on the branching fractions at 95\% CL are given by
\begin{equation} \label{eq:fom_ul}
\mathcal{B}(\tau\to\ell^{-}\alpha) = \frac{n_{\mathrm{sig}}^{95}}{2\epsilon N_{\tau\tau}},
\end{equation}  
where $\epsilon$ is the signal detection efficiency for a given $\alpha$ mass.

We estimate systematic uncertainties in the measured branching fractions of $\tau\to\ell\alpha$ decays arising from various sources.
The uncertainties in track reconstruction efficiencies are estimated with partially reconstructed $D^{*+}\rightarrow D^{0}\pi^{+}$, $D^{0}\rightarrow K^{0}_{S}\pi^{+}\pi^{-}$ events.
A systematic uncertainty of $0.35\%$ is assigned per track. Considering both two-track and four-track candidates, a total uncertainty of $1.1\%$ is estimated for this analysis.
The uncertainties in the lepton identification efficiencies are estimated with $J/\psi\to\ell^{+}\ell^{-}$~($\ell=e,\mu$) events and rates to misidentify a pion as a lepton are estimated with $\tau^{-}\to\pi^{-}\pi^{+}\pi^{-}\nu_{\tau}$ events.
The difference of the identification and misidentification efficiencies between data and MC simulation is considered and they are estimated to be $2.3$--$2.6\%$ for the $\tau^{-}\rightarrow e^{-}\alpha$ search and $1.1$--$1.6\%$ for the $\tau^{-}\rightarrow\mu^{-}\alpha$ search, where the assigned uncertainties for different $\alpha$ masses lie within the presented ranges.
The uncertainties in trigger efficiencies are investigated by MC samples with a trigger simulation. They are obtained by comparing the efficiencies with and without the trigger simulation, and estimated to be $4.2$--$5.2\%$ for the $\tau^{-}\rightarrow e^{-}\alpha$ search and $4.1$--$5.8\%$ for the $\tau^{-}\rightarrow\mu^{-}\alpha$ search. 
These sources of systematic uncertainties are related to the signal detection efficiencies. The uncertainty in the integrated luminosity is $1.4\%$ and in the  $e^{+}e^{-}\to\tau^{+}\tau^{-}$ cross section is 0.3\%~\cite{Ntautau}. These sources of systematic uncertainties are related to $N_{\tau\tau}$.
The total systematic uncertainty related to the quantity $1/(2\epsilon N_{\tau\tau})$ is estimated by summing in quadrature: $5.1$--$6.0$\% for the $\tau^{-}\rightarrow e^{-}\alpha$ search and $4.7$--$6.2\%$ for the $\tau^{-}\rightarrow\mu^{-}\alpha$ search.
We convolve the distribution of $n_{\mathrm{sig}}$ obtained from the likelihood fit to pseudoexperiments with a Gaussian distribution with a relative standard deviation corresponding to the total systematic uncertainty.
Additionally, the uncertainty due to PDF modeling is evaluated by varying several systematic sources such as lepton identification, trigger efficiencies and cross-sections of background components.
The number of signal events obtained from the fit is checked when each systematic source is varied. We assume the variations not to be correlated with each other, and thus the quadratic sum of the variation is considered as the upper limit.

\section{Results}
\begin{figure}[htbp]
 \begin{minipage}{0.45\hsize}
  \begin{center}
   \includegraphics[width=70mm]{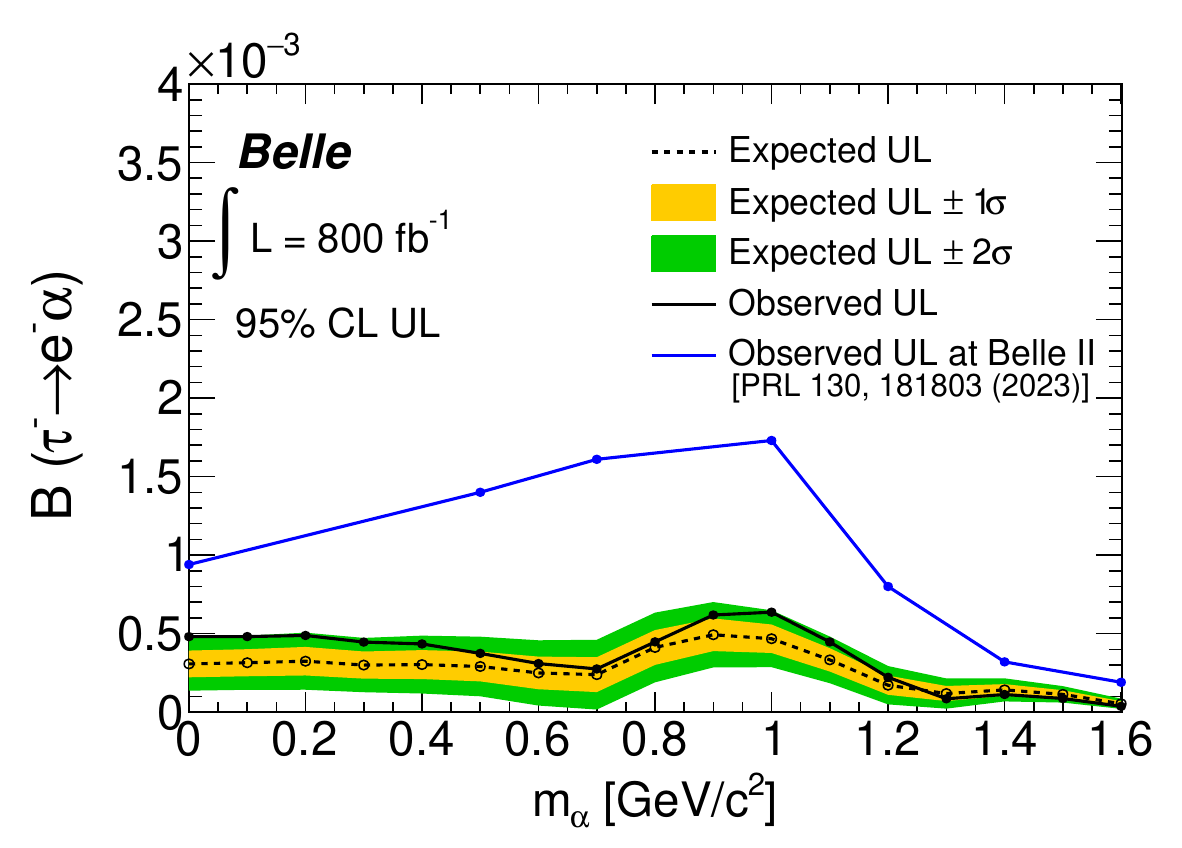}
   \hspace{1.5 cm} {(a) $\tau^{-}\rightarrow e^{-}\alpha$}
  \end{center}
 \end{minipage}
 \begin{minipage}{0.45\hsize}
  \begin{center}
   \includegraphics[width=70mm]{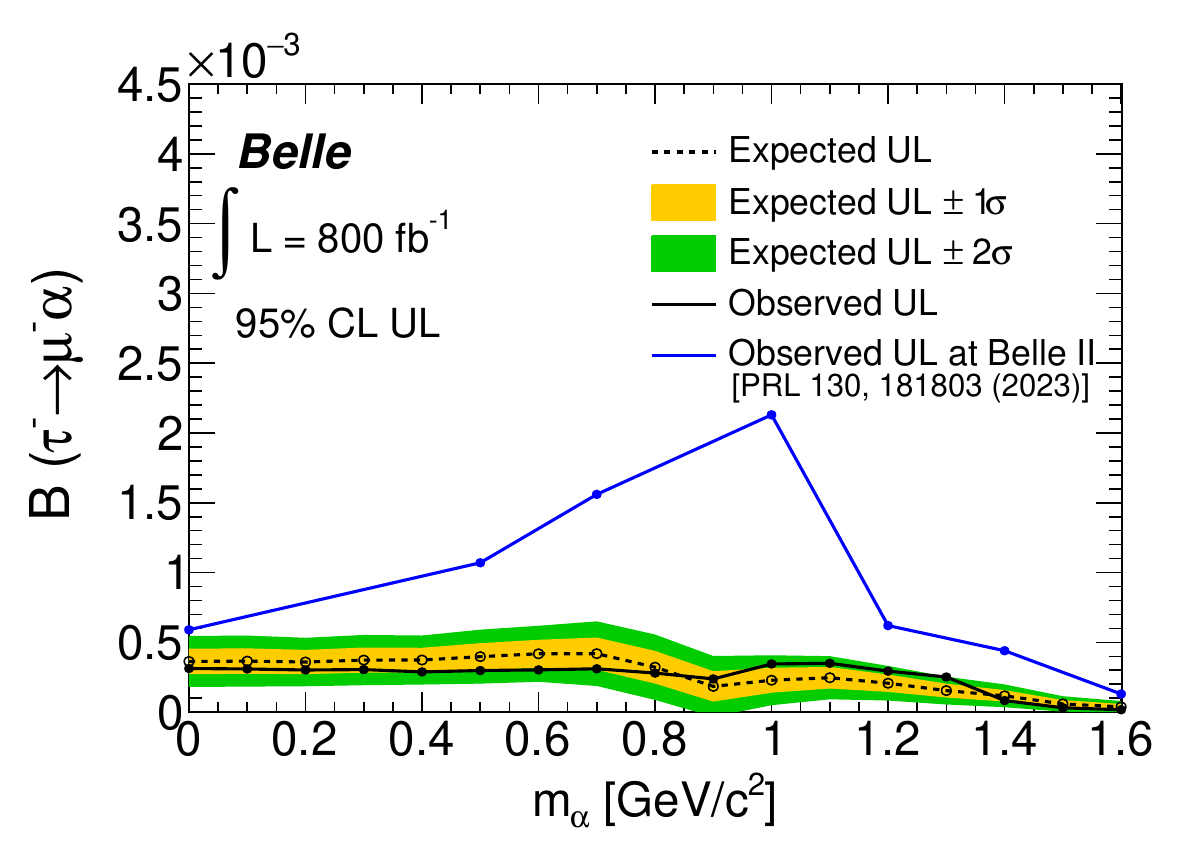}
    \hspace{1.5 cm} {(b) $\tau^{-}\rightarrow \mu^{-}\alpha$}
  \end{center}
 \end{minipage}
\caption{ \label{fig:UL95} 
  Upper limit at the 95\% CL on the branching fraction of $\mathcal{B}(\tau^{-}\to e^{-}\alpha)$~(a) and $\mathcal{B}(\tau^{-}\to \mu^{-}\alpha)$~(b) as a function of the $\alpha$ mass.
  The dashed~(solid) lines are expected~(observed) upper limits. The blue line shows the observed upper limits at Belle II~\cite{BelleII}.
}
\end{figure}

We observe no significant excess of signal events over the background prediction in data and thus, evaluate upper limits on the branching fractions.
Figure~\ref{fig:UL95} shows the upper limits at 95\% CL on the branching fractions of $\mathcal{B}(\tau^{-}\to e^{-}\alpha)$ and $\mathcal{B}(\tau^{-}\to \mu^{-}\alpha)$ as a function of the $\alpha$ mass.

The observed upper limits on $\mathcal{B}(\tau^{-}\rightarrow\ell^{-}\alpha)$ at the $95\%$ CL are calculated:
$\mathcal{B}(\tau^{-}\rightarrow e^{-}\alpha)<(0.4-6.4) \times 10^{-4}$, $\mathcal{B}(\tau^{-}\rightarrow\mu^{-}\alpha)<(0.2-3.5) \times 10^{-4}$ at 95\% CL.
The ranges indicate the dependence on the $\alpha$ mass. Our results are the most stringent limits to date.
We also provide an appendix containing the observed upper limits at the 90\% CL obtained in a similar way.

\section{Conclusions}
In this paper, a search for the lepton-flavor violating decays, $\tau^{-}\to\ell^{-}\alpha$~($\ell=e,\mu$), based on a sample of $736\times10^{6}$ tau pairs recorded at the Belle experiment is reported.
We observe no significant excess of signal events in data over background and set upper limits on the branching fraction $\mathcal{B}(\tau^{-}\to\ell^{-}\alpha)$ at the 95\% CL.
These results are the most stringent limits on tau decays with an invisible spin-0 particle.

\acknowledgments
This work, based on data collected using the Belle detector, which was
operated until June 2010, was supported by 
the Ministry of Education, Culture, Sports, Science, and
Technology (MEXT) of Japan, the Japan Society for the 
Promotion of Science (JSPS), and the Tau-Lepton Physics 
Research Center of Nagoya University; 
the Australian Research Council including grants
DP210101900, % Urquijo
DP210102831, % Sevior
DE220100462, % Hsu
LE210100098, % Infrastructure
LE230100085; % Infrastructure
Austrian Federal Ministry of Education, Science and Research and
Austrian Science Fund (FWF) No.~P~31361-N36;
National Key R\&D Program of China under Contract No.~2022YFA1601903,
National Natural Science Foundation of China and research grants
No.~11575017,
No.~11761141009, 
No.~11705209, 
No.~11975076, 
No.~12135005, 
No.~12150004, 
No.~12161141008, 
No.~12175041, 
and
No.~12475093,
and Shandong Provincial Natural Science Foundation Project ZR2022JQ02;
the Czech Science Foundation Grant No. 22-18469S;
Horizon 2020 ERC Advanced Grant No.~884719 and ERC Starting Grant No.~947006 ``InterLeptons'' (European Union);
the Carl Zeiss Foundation, the Deutsche Forschungsgemeinschaft, the
Excellence Cluster Universe, and the VolkswagenStiftung;
the Department of Atomic Energy (Project Identification No. RTI 4002), the Department of Science and Technology of India,
and the UPES (India) SEED finding programs Nos. UPES/R\&D-SEED-INFRA/17052023/01 and UPES/R\&D-SOE/20062022/06; 
the Istituto Nazionale di Fisica Nucleare of Italy; 
National Research Foundation (NRF) of Korea Grant
Nos.~2016R1-D1A1B-02012900, 
2018R1-A6A1A-06024970, 
2021R1-A6A1A-03043957, 
2021R1-F1A-1060423, 
2021R1-F1A-1064008, 
2022R1-A2C-1003993, 
2022R1-A2C-1092335, 
RS-2022-00197659, 
RS-2023-00208693; 
Radiation Science Research Institute, Foreign Large-size Research Facility Application Supporting project, the Global Science Experimental Data Hub Center, the Korea Institute of Science and Technology Information (K24L2M1C4) and KREONET/GLORIAD; 
the Polish Ministry of Science and Higher Education and 
the National Science Center;
the Ministry of Science and Higher Education of the Russian Federation
and the HSE University Basic Research Program, Moscow; % from 15.04.2021
University of Tabuk research grants
S-1440-0321, S-0256-1438, and S-0280-1439 (Saudi Arabia);
the Slovenian Research Agency Grant Nos. J1-9124 and P1-0135;
Ikerbasque, Basque Foundation for Science, and the State Agency for Research
of the Spanish Ministry of Science and Innovation through Grant No. PID2022-136510NB-C33 (Spain);
the Swiss National Science Foundation; 
the Ministry of Education and the National Science and Technology Council of Taiwan;
and the United States Department of Energy and the National Science Foundation.
These acknowledgements are not to be interpreted as an endorsement of any
statement made by any of our institutes, funding agencies, governments, or
their representatives.
We thank the KEKB group for the excellent operation of the
accelerator; the KEK cryogenics group for the efficient
operation of the solenoid; and the KEK computer group and the Pacific Northwest National
Laboratory (PNNL) Environmental Molecular Sciences Laboratory (EMSL)
computing group for strong computing support; and the National
Institute of Informatics, and Science Information NETwork 6 (SINET6) for
valuable network support.

  \section*{Appendix}

\begin{figure}[htbp]
 \begin{minipage}{0.45\hsize}
  \begin{center}
   \includegraphics[width=70mm]{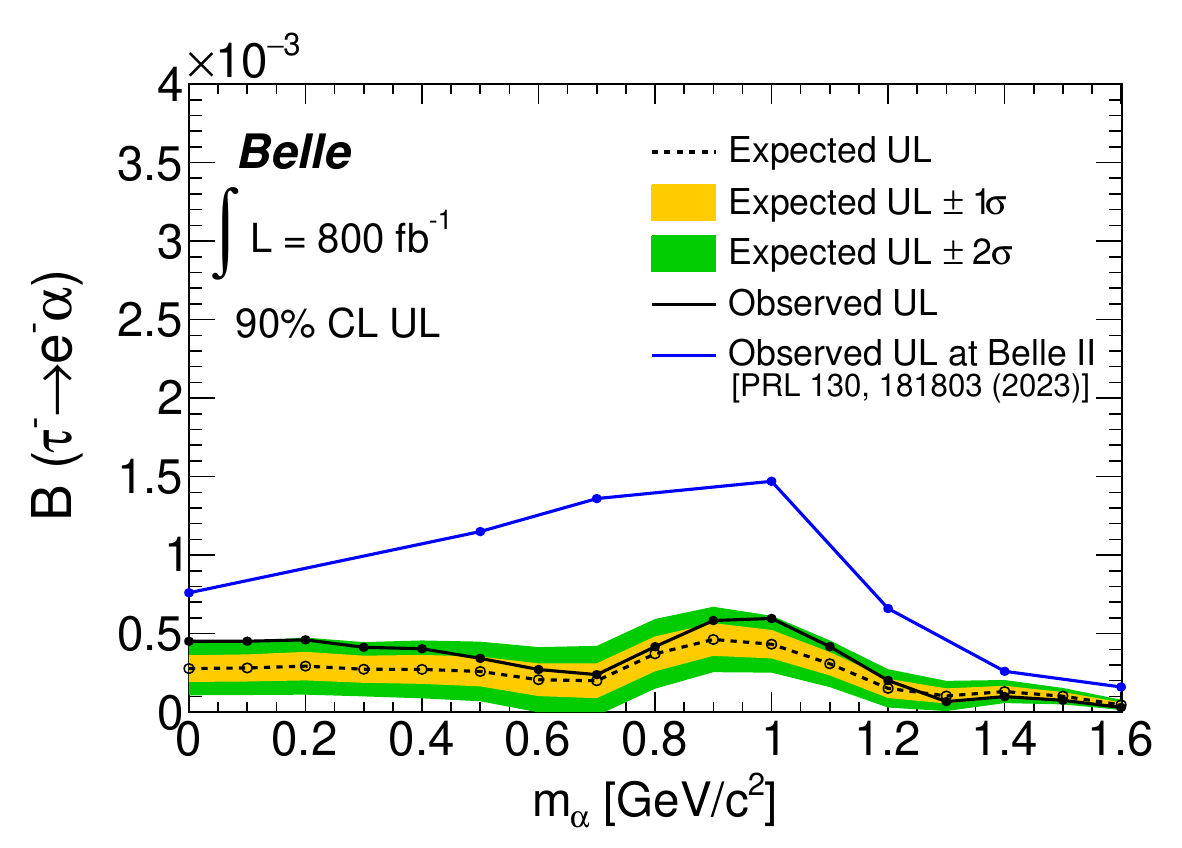} \\
   \hspace{1.5 cm} {(a) $\tau^{-}\rightarrow e^{-}\alpha$}
  \end{center}
 \end{minipage}
 \begin{minipage}{0.45\hsize}
  \begin{center}
   \includegraphics[width=70mm]{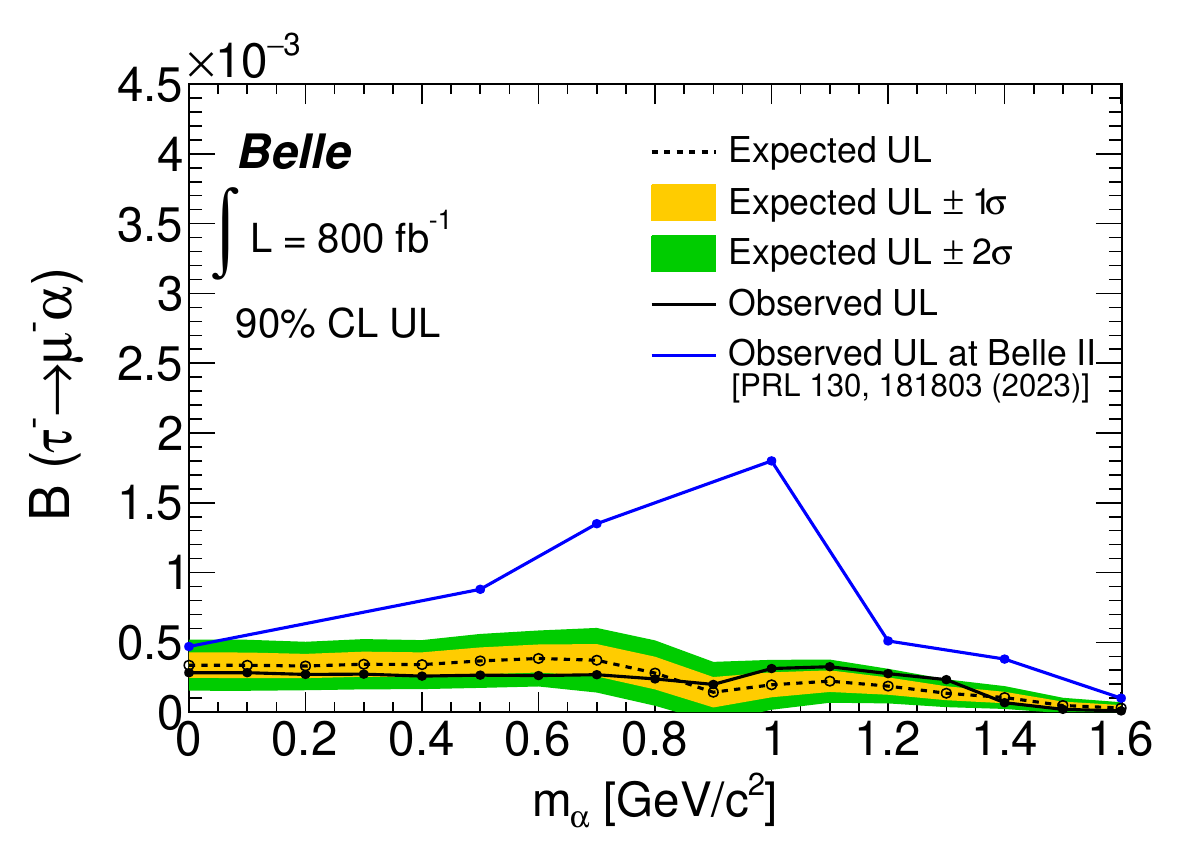}
    \hspace{1.5 cm} {(b) $\tau^{-}\rightarrow \mu^{-}\alpha$}
  \end{center}
 \end{minipage}
\caption{ \label{fig:UL90} 
  Upper limit at the 90\% CL on the branching fraction of $\mathcal{B}(\tau^{-}\to e^{-}\alpha)$~(a) and $\mathcal{B}(\tau^{-}\to \mu^{-}\alpha)$~(b) as a function of the $\alpha$ mass.
  The dashed~(solid) lines are expected~(observed) upper limits. The blue line shows the observed upper limits at Belle II~\cite{BelleII}.
}
\end{figure}

Figure~\ref{fig:UL90} shows the upper limits at the 90\% CL on the branching fractions of $\mathcal{B}(\tau^{-}\rightarrow e^{-}\alpha)$ and $\mathcal{B}(\tau^{-}\rightarrow\mu^{-}\alpha)$ as a function of the $\alpha$ mass.
The observed upper limits on $\mathcal{B}(\tau^{-}\rightarrow\ell^{-}\alpha)$ at the $90\%$ CL are calculated:
$\mathcal{B}(\tau^{-}\rightarrow e^{-}\alpha)<(0.3-6.0) \times 10^{-4}$, $\mathcal{B}(\tau^{-}\rightarrow\mu^{-}\alpha)<(0.1-3.3) \times 10^{-4}$ at 90\% CL. The ranges indicate the dependence on the $\alpha$ mass.

\begin{figure}[]
 \begin{minipage}{0.45\hsize}
  \begin{center}
   \includegraphics[width=70mm]{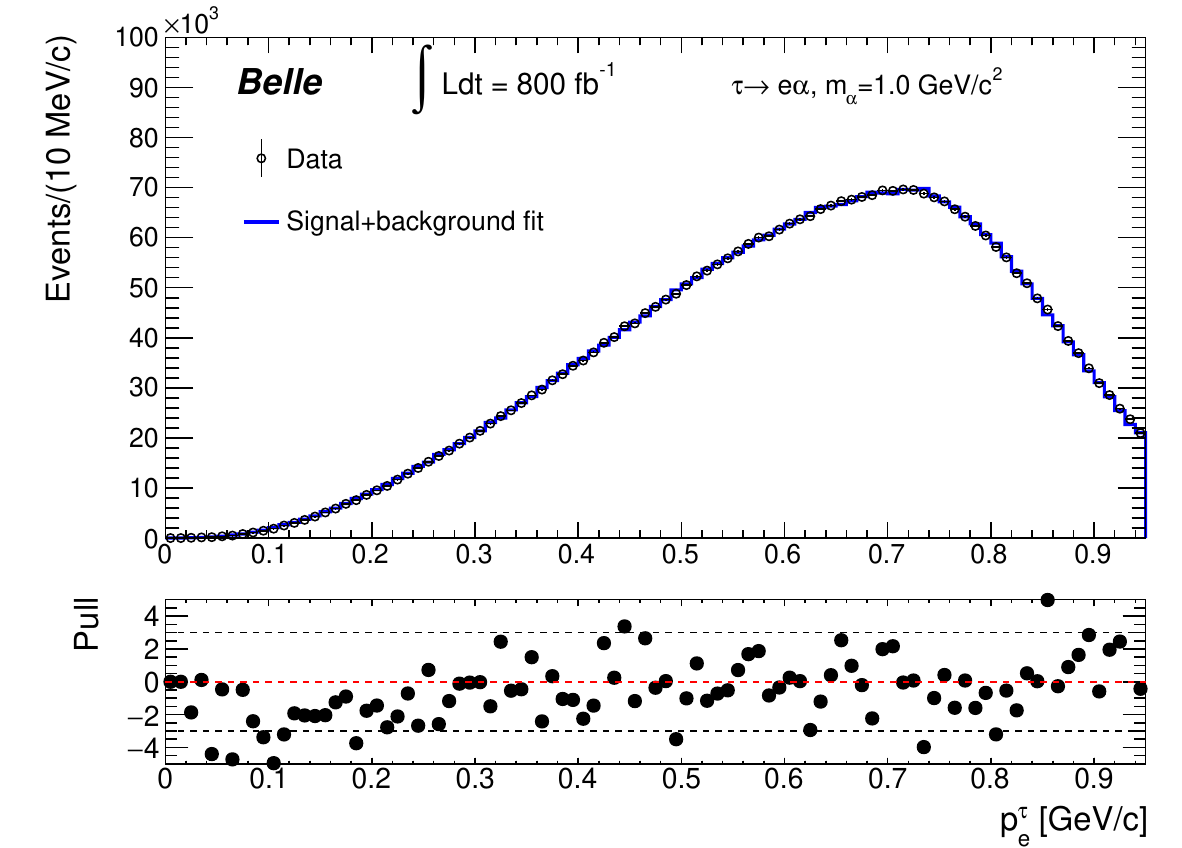} \\
   \hspace{1.5 cm} {(a) $\tau^{-}\rightarrow e^{-}\alpha$}
  \end{center}
 \end{minipage}
 \begin{minipage}{0.45\hsize}
  \begin{center}
   \includegraphics[width=70mm]{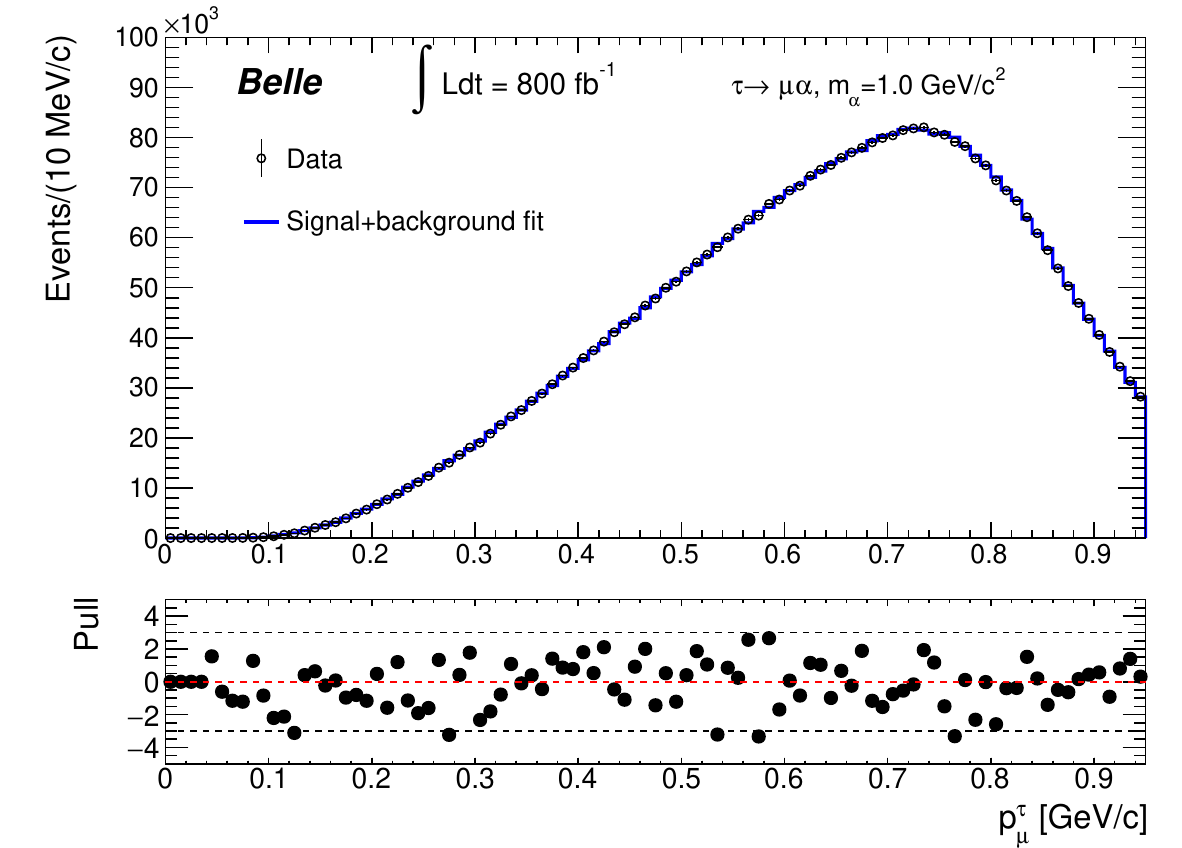}
    \hspace{1.5 cm} {(b) $\tau^{-}\rightarrow \mu^{-}\alpha$}
  \end{center}
 \end{minipage}
\caption{ \label{fig:post-fit} 
    Examples of the fit to the distributions of $p_{\ell}^{\tau}$ for the $\tau^{-}\to e^{-}\alpha$ candidates with $m_\alpha=1.0$~GeV/$c^{2}$ and the $\tau^{-}\to \mu^{-}\alpha$ candidates with $m_\alpha=1.0$~GeV/$c^{2}$ are shown. The blue lines show the overall fit result. The bottom figures show the pull distributions defined as the differences between the data and fit results divided by the statistical uncertainty of the data.}
\end{figure}

The detection efficiencies, systematic uncertainties, fitted signal yields with their statistical uncertainties and upper limits are summarized in Tables~\ref{tab:result_ealpha} and \ref{tab:result_malpha}.
The $\alpha$ mass with the highest signal strength is $1.0$~GeV/$c^{2}$ for both the $\tau^{-}\to e^{-}\alpha$ and $\tau^{-}\to\mu^{-}\alpha$ searches. The extracted signal yields are $3224.0\pm1331.6$ and $2368.1\pm2394.9$ for the $\tau^{-}\to e^{-}\alpha$ and $\tau^{-}\to\mu^{-}\alpha$ searches, respectively.
The fits to the distributions of $p_{\ell}^{\tau}$ with the signal PDFs using $m_{\alpha}=1.0$~GeV/$c^{2}$ are shown in Figure~\ref{fig:post-fit}.

\begin{table}[htbp]
  \begin{center}
      \caption{ \label{tab:result_ealpha} Signal detection efficiencies~($\epsilon_{\mathrm{sig}}$), systematic uncertainties related to the efficiencies~($\sigma_{\mathrm{syst}}$), fitted signal yields with their statistical uncertainties~($n_{\mathrm{sig}}$), measured branching fractions with their statistical uncertainties~($\mathcal{B}(10^{-4})$), expected upper limits at 90\% and 95\% CL~($\mathcal{B}^{\mathrm{Exp}}$), Observed upper limits at the 90\% and 95\% CL~($\mathcal{B}^{\mathrm{Obs}}$) for the $\tau^{-}\to e^{-}\alpha$ events with various $\alpha$ mass~($m_{\alpha}$).}
      \scalebox{0.73}[0.73]{
       \begin{tabular}{ c  c  c  c  c  c  c  c  c } \hline
         %m$_{\alpha}$ [GeV/$c^{2}$] & $\epsilon_{\mathrm{sig}}$ [\%]  & $\sigma_{\mathrm{syst}}$ [\%] & $n_{\mathrm{sig}}$ & $\mathcal{B}^{\mathrm{Exp}}$~(10$^{-4}$) at 95\% CL & $\mathcal{B}^{\mathrm{Obs}}$~(10$^{-4}$) at 95\% CL & $\mathcal{B}^{\mathrm{Exp}}$~(10$^{-4}$) at 90\% CL & $\mathcal{B}^{\mathrm{Obs}}$~(10$^{-4}$) at 90\% CL \\ \hline \hline
         m$_{\alpha}$ [GeV/$c^{2}$] & $\epsilon_{\mathrm{sig}}$ [\%]  & $\sigma_{\mathrm{syst}}$ [\%] & $n_{\mathrm{sig}}$ & $\mathcal{B}(10^{-4})$ & \makecell{$\mathcal{B}^{\mathrm{Exp}}$~(10$^{-4}$) \\ at 95\% CL} & \makecell{$\mathcal{B}^{\mathrm{Obs}}$~(10$^{-4}$) \\ at 95\% CL} & \makecell{$\mathcal{B}^{\mathrm{Exp}}$~(10$^{-4}$) \\ at 90\% CL} & \makecell{$\mathcal{B}^{\mathrm{Obs}}$~(10$^{-4}$) \\ at 90\% CL} \\ \hline \hline
         
        0.0 & 0.9 & 5.7 & \multicolumn{1}{r}{2260.3 $\pm$ 852.2} & 1.7 $\pm$ 0.6 & 3.1 & 4.8 & 2.8 & 4.5 \\ 
        0.1 & 0.9 & 5.8 & \multicolumn{1}{r}{2246.1 $\pm$ 957.8} & 1.7 $\pm$ 0.7 & 3.1 & 4.8 & 2.8 & 4.5 \\ 
        0.2 & 1.0 & 5.8 & \multicolumn{1}{r}{2327.2 $\pm$ 1064.3} & 1.6 $\pm$ 0.7 & 3.2 & 4.9 & 2.9 & 4.6 \\
        0.3 & 1.0 & 5.9 & \multicolumn{1}{r}{2041.4 $\pm$ 989.5} & 1.4 $\pm$ 0.7 & 3.0 & 4.5 & 2.7 & 4.1 \\ 
        0.4 & 1.1 & 5.9 & \multicolumn{1}{r}{2090.0 $\pm$ 1176.0} & 1.3 $\pm$ 0.7 & 3.0 & 4.3 & 2.7 & 4.0 \\
        0.5 & 1.2 & 5.9 & \multicolumn{1}{r}{1392.1 $\pm$ 1203.1} & 0.8 $\pm$ 0.7 & 2.9 & 3.7 & 2.6 & 3.4 \\
        0.6 & 1.3 & 5.9 & \multicolumn{1}{r}{1115.5 $\pm$ 1278.9} & 0.6 $\pm$ 0.7 & 2.5 & 3.1 & 2.1 & 2.7 \\
        0.7 & 1.3 & 5.9 & \multicolumn{1}{r}{1197.0 $\pm$ 1347.4} & 0.6 $\pm$ 0.7 & 2.4 & 2.7 & 2.0 & 2.4 \\
        0.8 & 1.3 & 5.9 & \multicolumn{1}{r}{819.4 $\pm$ 1396.8} & 0.4 $\pm$ 0.7 & 4.1 & 4.5 & 3.7 & 4.2 \\ 
        0.9 & 1.4 & 5.8 & \multicolumn{1}{r}{2436.7 $\pm$ 1398.5} & 1.2 $\pm$ 0.7 & 4.9 & 6.2 & 4.6 & 5.8 \\
        1.0 & 1.4 & 5.9 & \multicolumn{1}{r}{3224.0 $\pm$ 1331.6} & 1.6 $\pm$ 0.6 & 4.7 & 6.4 & 4.3 & 6.0 \\
        1.1 & 1.4 & 5.7 & \multicolumn{1}{r}{2178.4 $\pm$ 1194.5} & 1.1 $\pm$ 0.6 & 2.8 & 4.5 & 2.7 & 4.2 \\
        1.2 & 1.3 & 5.6 & \multicolumn{1}{r}{1016.9 $\pm$ 975.9} & 0.5 $\pm$ 0.5 & 1.7 & 2.2 & 1.5 & 2.0 \\ 
        1.3 & 1.3 & 5.5 & \multicolumn{1}{r}{$-$681.3 $\pm$ 778.4} & $-$0.4 $\pm$ 0.4 & 1.2 & 0.8 & 1.0 & 0.7 \\ 
        1.4 & 1.3 & 5.1 & \multicolumn{1}{r}{$-$583.4 $\pm$ 620.5} & $-$0.3 $\pm$ 0.3 & 1.4 & 1.1 & 1.3 & 1.0 \\ 
        1.5 & 1.2 & 5.3 & \multicolumn{1}{r}{$-$451.9 $\pm$ 437.6} & $-$0.3 $\pm$ 0.2 & 1.1 & 0.9 & 1.0 & 0.8 \\ 
        1.6 & 1.0 & 6.0 & \multicolumn{1}{r}{$-$229.5 $\pm$ 224.0} & $-$0.2 $\pm$ 0.2 & 0.5 & 0.4 & 0.5 & 0.3 \\ \hline \hline
      \end{tabular}
    }
  \end{center}
\end{table}

\begin{table}[htbp]
  \begin{center}
    \caption{ \label{tab:result_malpha} Signal detection efficiencies~($\epsilon_{\mathrm{sig}}$), systematic uncertainties related to the efficiencies~($\sigma_{\mathrm{syst}}$), fitted signal yields with their statistical uncertainties~($n_{\mathrm{sig}}$), measured branching fractions with their statistical uncertainties~($\mathcal{B}(10^{-4})$), expected upper limits at 90\% and 95\% CL~($\mathcal{B}^{\mathrm{Exp}}$), observed upper limits at the 90\% and 95\% CL~($\mathcal{B}^{\mathrm{Obs}}$) for the $\tau^{-}\to \mu^{-}\alpha$ events with various $\alpha$ mass~($m_{\alpha}$).}    
      \scalebox{0.73}[0.73]{
      \begin{tabular}{ c  c  c  c  c  c  c  c  c} \hline
        %m$_{\alpha}$ [GeV/$c^{2}$] & $\epsilon_{\mathrm{sig}}$ [\%]  & $\sigma_{\mathrm{syst}}$ [\%] & $n_{\mathrm{sig}}$ & $\mathcal{B}_{95}^{\mathrm{Exp}}$~(10$^{-4}$) & $\mathcal{B}_{95}^{\mathrm{Obs}}$~(10$^{-4}$) & $\mathcal{B}_{90}^{\mathrm{Exp}}$~(10$^{-4}$) & $\mathcal{B}_{90}^{\mathrm{Obs}}$~(10$^{-4}$) \\ \hline \hline
        m$_{\alpha}$ [GeV/$c^{2}$] & $\epsilon_{\mathrm{sig}}$ [\%]  & $\sigma_{\mathrm{syst}}$ [\%] & $n_{\mathrm{sig}}$ & $\mathcal{B}(10^{-4})$  & \makecell{$\mathcal{B}^{\mathrm{Exp}}$~(10$^{-4}$) \\ at 95\% CL} & \makecell{$\mathcal{B}^{\mathrm{Obs}}$~(10$^{-4}$) \\ at 95\% CL} & \makecell{$\mathcal{B}^{\mathrm{Exp}}$~(10$^{-4}$) \\ at 90\% CL} & \makecell{$\mathcal{B}^{\mathrm{Obs}}$~(10$^{-4}$) \\ at 90\% CL} \\ \hline \hline        
        0.0 & 1.0 & 6.2 & \multicolumn{1}{r}{$-$764.3 $\pm$ 723.3}   & $-$0.5 $\pm$ 0.5 & 3.6 & 3.1 & 3.4 & 2.8 \\ 
        0.1 & 1.0 & 6.2 & \multicolumn{1}{r}{$-$817.0 $\pm$ 1191.5}  & $-$0.6 $\pm$ 0.8 & 3.7 & 3.1 & 3.4 & 2.8 \\
        0.2 & 1.1 & 6.0 & \multicolumn{1}{r}{$-$919.8 $\pm$ 1131.8}  & $-$0.6 $\pm$ 0.7 & 3.6 & 3.0 & 3.3 & 2.7 \\
        0.3 & 1.1 & 5.9 & \multicolumn{1}{r}{$-$1203.0 $\pm$ 887.9}  & $-$0.7 $\pm$ 0.5 & 3.7 & 3.0 & 3.4 & 2.7 \\
        0.4 & 1.2 & 5.7 & \multicolumn{1}{r}{$-$1479.6 $\pm$ 1336.6} & $-$0.8 $\pm$ 0.8 & 3.7 & 2.9 & 3.4 & 2.6 \\
        0.5 & 1.3 & 5.7 & \multicolumn{1}{r}{$-$1951.0 $\pm$ 1600.2} & $-$1.0 $\pm$ 0.8 & 4.0 & 3.0 & 3.7 & 2.6 \\
        0.6 & 1.4 & 5.8 & \multicolumn{1}{r}{$-$2516.4 $\pm$ 1819.1} & $-$1.2 $\pm$ 0.9 & 4.2 & 3.0 & 3.8 & 2.6 \\
        0.7 & 1.4 & 5.8 & \multicolumn{1}{r}{$-$2236.7 $\pm$ 1977.6} & $-$1.1 $\pm$ 1.0 & 4.2 & 3.1 & 3.7 & 2.7 \\
        0.8 & 1.5 & 5.8 & \multicolumn{1}{r}{$-$842.6 $\pm$ 1673.9}  & $-$0.4 $\pm$ 0.8 & 3.2 & 2.8 & 2.8 & 2.4 \\
        0.9 & 1.5 & 5.8 & \multicolumn{1}{r}{1152.3 $\pm$ 1744.5}    & 0.5 $\pm$ 0.8 & 1.8 & 2.4 & 1.4 & 2.0 \\
        1.0 & 1.5 & 5.5 & \multicolumn{1}{r}{2368.1 $\pm$ 2394.9}    & 1.1 $\pm$ 1.1 & 2.3 & 3.4 & 2.0 & 3.1 \\
        1.1 & 1.4 & 5.5 & \multicolumn{1}{r}{2026.0 $\pm$ 1104.7}    & 1.0 $\pm$ 0.5 & 2.5 & 3.5 & 2.2 & 3.3 \\
        1.2 & 1.3 & 5.3 & \multicolumn{1}{r}{1699.5 $\pm$ 985.5}     & 0.9 $\pm$ 0.5 & 2.1 & 2.9 & 1.9 & 2.8 \\ 
        1.3 & 1.3 & 5.1 & \multicolumn{1}{r}{1747.6 $\pm$ 1679.9}    & 0.9 $\pm$ 0.9 & 1.5 & 2.5 & 1.4 & 2.3 \\
        1.4 & 1.1 & 4.7 & \multicolumn{1}{r}{$-$555.1 $\pm$ 810.7}   & $-$0.3 $\pm$ 0.5 & 1.2 & 0.8 & 1.0 & 0.7 \\ 
        1.5 & 0.9 & 4.8 & \multicolumn{1}{r}{$-$348.3 $\pm$ 311.4}   & $-$0.3 $\pm$ 0.2 & 0.6 & 0.3 & 0.5 & 0.2 \\
        1.6 & 0.3 & 4.8 & \multicolumn{1}{r}{$-$99.3 $\pm$ 95.0}     & $-$0.2 $\pm$ 0.2 & 0.4 & 0.2 & 0.3 & 0.1 \\ \hline \hline
        \end{tabular}
      }
  \end{center}
\end{table}

\end{document}